\documentclass[a4paper,notitlepage,twocolumn,superscriptaddress,longbibliography,nofootinbib]{revtex4-1}
\usepackage {amsmath}
\usepackage{amssymb}
 \usepackage{amsfonts}

\begin{document}

\title{One-loop Correction to the AdS/BCFT Partition Function in the Three Dimensional Pure Gravity}

\author{Yu-ki Suzuki}
\email{yu-ki.suzuki@yukawa.kyoto-u.ac.jp}
\affiliation{Center for Gravitational Physics, Yukawa Institute for Theoretical Physics,\\ 
Kitashirakawa Oiwakecho, Sakyo-ku, Kyoto University, Kyoto 606-8502, Japan}

\begin{abstract}
We calculate the tree-level partition function of an Euclidean BTZ black
hole in the presence of an end of the world branes (ETW branes)
and the one-loop partition function of an Euclidean thermal $AdS_{3}$
in the presence of an ETW brane. 
At the tree level, our results match with the previous
ones for static BTZ black holes. Interestingly, at the one-loop level, our results
contain novel terms, which reflect the existence of an ETW brane.
The ETW brane has a consistent profile for a static thermal AdS and in this case we 
showed that the spectrum obtained from the one-loop partition function takes a physically sensible form.
\end{abstract}

\maketitle


\section{Introduction}

The one-loop partition function in gravity was calculated
almost half a century ago in the famous work \citep{tHooft:1974toh}.
In the context of the AdS/CFT correspondence \citep{Maldacena:1997re},
there have been several works toward deriving the full quantum gravity partition
function of pure gravity at a finite temperature \citep{Giombi:2008vd,Maloney:2007ud,Yin:2007gv}.
At the one-loop level, they deduced the result from the CFT path
integral calculation \citep{Maloney:2007ud}, and later this was directly proved
using the heat kernel method \citep{Giombi:2008vd}, summarized into the formula:
\begin{equation}
Z_{gravity}=\prod^\infty_{m=2}\frac{1}{\left|1-q^{m}\right|^{2}},\label{eq:-2}
\end{equation}
where $q$ describes the moduli of the boundary torus.

Our goal is to extend this calculation to the AdS/BCFT case.
The AdS/BCFT correspondence is an extension of the AdS/CFT correspondence to the case where a CFT lives on a manifold with boundaries (called boundary conformal field theory, or BCFT)
\citep{Karch:2000gx,Takayanagi:2011zk,Fujita:2011fp,Nozaki:2012qd}.
The basic idea of the AdS/BCFT is to extend the boundary of the manifold where the CFT is defined, to a codimension one surface in the AdS, called the end of the world brane (ETW brane). The gravity in the spacetime surrounded by this ETW brane, provides the gravity dual of the BCFT. In this paper, first, we will calculate the tree-level partition
function of a rotating Euclidean BTZ black hole in the presence of ETW branes with an arbitrary tension, which is summarized as the result:
\begin{equation}
\thickmuskip=0mu
\medmuskip=0mu
\thinmuskip=0mu Z_{BTZ-tree}=\exp\left[\frac{\pi R_{+}}{8G}+\frac{1}{8G}\left(\log\left(\frac{1+T_{\pi}}{1-T_{\pi}}\right)-\log\left(\frac{1+T_{0}}{1-T_{0}}\right)\right)\right].
\end{equation}
In the static (i.e. non-rotating) case, this matches with the previous calculation \citep{Fujita:2011fp}.
Secondly, we will calculate the one-loop partition function of a thermal
AdS$_{3}$ in the presence of a tensionless ETW brane  with the Neumann boundary condition (for earlier discussions of the boundary conditions in gravity refer to \citep{Moss:1996ip,vanNieuwenhuizen:2005kg}). Our final result is given by
\begin{equation}
\thickmuskip=0mu
\medmuskip=0mu
\thinmuskip=0mu Z_{gravity}=\prod_{m=2}^{\infty}\frac{1}{\left|1-q^{m}\right|}\cdot \prod_{l=0}^{\infty}\frac{\sqrt{1-q^{l+2}\overline{q}^{l+1}}\sqrt{1-q^{l+1}\overline{q}^{l+2}}}{\sqrt{1-q^{l+2}\overline{q}^{l}}\sqrt{1-q^{l}\overline{q}^{l+2}}}.\label{eq:-1}
\end{equation}
There are two contributions: the first term corresponds to a square root of
the original result, which comes from the bulk modes, and the second term is the new effect due to the ETW brane. The second term 
represents the contributions from a massive ghost vector field, due to the BRST quantization of gravity, and the massless spin-2 field. However, as we consider later, if we set $q=\overline{q}$ to have a consistent configuration of ETW brane,
then we have a physically sensible partition function with non-negative coefficients in the $q$ expansion like (\ref{eq:-2}). 

One might wonder if we can impose the Dirichlet boundary condition on the ETW brane \citep{Miao:2018qkc}. At the tree level, both the
Neumann and the Dirichlet boundary condition lead to an identical partition function for a tensionless ETW brane. On the other hand, at the one-loop level, the resulting partition function for the Dirichlet boundary condition is different from that for the Neumann one. However, as we will see later, we find that the Dirichlet boundary condition is in tension with the BRST invariance.

Finally, we will discuss summing over the modular transformation. The boundary
CFT lives on the conformal boundary of a half of the solid torus \citep{Kraus:2006wn} in our AdS/BCFT setup. 
We expect that it still admits modularity
even if we insert branes because from the boundary torus perspective
it still admits modular invariance and then we consider how to locate
the position of the brane in the bulk. To derive the full partition
function, we must sum over the $SL(2,\mathbb{Z})$ transformation
of the thermal AdS contribution \citep{Dijkgraaf:2000fq,Maloney:2007ud}.

This paper is organized as follows. In section 2 we directly calculate
the partition function of a rotating BTZ black hole with ETW branes.
There are already earlier works on this calculation in the no boundary case \citep{Banados:1992gq,Carlip:1994gc,Gibbons:1976ue,Maldacena:1998bw}.
In section 3, we review the heat kernel method for calculating a  1-loop
partition function \citep{David:2009xg,Gopakumar:2011qs,Mann:1996ze,Vassilevich:2003xt} and then we apply this to the computation of a 
1-loop partition function of the scalar, the
vector, and the spin-2 fields using a method of images in the AdS/BCFT. 
In section 4, we discuss the physical interpretation of the partition function
and the consistency of the boundary conditions. In section 5 we discuss
our conclusions and present some future directions. In appendix
A we give the detailed calculation of derivatives of chordal
distance $u$. In appendix B, we present the calculation of a partition function in the Dirichlet case.

\section{A tree-Level Partition Function of a BTZ black hole with ETW branes}

\subsection{BTZ with tensionless ETW branes}

Let us consider an Euclidean rotating BTZ black hole in three dimension
\citep{Banados:1992gq}. The metric is given by

\begin{eqnarray}
\thickmuskip=0mu
\medmuskip=0mu
\thinmuskip=0mu ds^{2}&=&\frac{(r^{2}-R_{+}^{2})(r^{2}+R_{-}^{2})}{r^{2}}dt^{2}+\frac{r^{2}}{(r^{2}-R_{+}^{2})(r^{2}+R_{-}^{2})}dr^{2}\nonumber \\
&+&r^{2}(d\phi-\frac{R_{+}R_{-}}{r^{2}}dt)^{2}.
\end{eqnarray}

Here $R_{-}$ is a real valued parameter and $R_{+}$ is the horizon radius. The absence of a conical singularity constrains the periodicity of time and rotational angle:
\begin{equation}
\left(t,\phi\right)\sim\left(t+\beta,\phi+\theta\right),
\end{equation}
where $\beta=\frac{2\pi R_{+}}{R_{+}^{2}+R_{-}^{2}}$ and $\theta=\frac{2\pi R_{-}}{R_{+}^{2}+R_{-}^{2}}$.
If we define new coordinate as $\phi^{\prime}\equiv\phi-\frac{\theta}{\beta}t$
, the periodicity can be recast as $\left(t,\phi^{\prime}\right)\sim\left(t+\beta,\phi^{\prime}\right)$
\citep{Carlip:1994gc}. We also rewrite the metric in terms of this
coordinate as
\begin{equation}
\thickmuskip=0mu
\medmuskip=0mu
\thinmuskip=0mug_{ab}=\left(\begin{array}{ccc}A & 0 & B\\
0 & C & 0\\
D & 0 & E
\end{array}\right),
\end{equation}
where 
\begin{eqnarray}
    A&=&\frac{(r^{2}-R_{+}^{2})(r^{2}+R_{-}^{2})}{r^{2}}+r^{2}\left(\frac{R_{-}}{R_{+}}-\frac{R_{+}R_{-}}{r^{2}}\right)^{2},\nonumber\\
    B&=&\frac{R_{-}}{R_{+}}r^{2}-R_{+}R_{-},\nonumber\\
    C&=&\frac{r^{2}}{(r^{2}-R_{+}^{2})(r^{2}+R_{-}^{2})},\nonumber\\
    D&=&\frac{R_{-}}{R_{+}}r^{2}-R_{+}R_{-},\nonumber\\
    E&=&r^{2}.
\end{eqnarray}

Note that det$(g_{ij})=r^{2}.$ To identify the location of an ETW brane we use the map to the Poincare coordinate:
\begin{equation}
\begin{array}{cc}
\eta=\left(\frac{r^{2}-R_{+}^{2}}{r^{2}+R_{-}^{2}}\right)^{\frac{1}{2}}\cos\left(\frac{R_{+}^{2}+R_{-}^{2}}{R_{+}}t+R_{-}\phi^{\prime}\right)\exp\left(R_{+}\phi^{\prime}\right),\\
x=\left(\frac{r^{2}-R_{+}^{2}}{r^{2}+R_{-}^{2}}\right)^{\frac{1}{2}}\sin\left(\frac{R_{+}^{2}+R_{-}^{2}}{R_{+}}t+R_{-}\phi^{\prime}\right)\exp\left(R_{+}\phi^{\prime}\right),\\
z=\left(\frac{R_{+}^{2}+R_{-}^{2}}{r^{2}+R_{-}^{2}}\right)^{\frac{1}{2}}\exp\left(R_{+}\phi^{\prime}\right),
\end{array}
\end{equation}
which leads to the familiar metric
\begin{equation}
ds^{2}=\frac{dz^{2}+dx^{2}+d\eta^{2}}{z^{2}}.
\end{equation}

In this coordinate the identification $(t,\phi^{\prime})\sim(t+\beta,\phi^{\prime})$
is trivial, but the identification $(t,\phi^{\prime})\sim(t,\phi^{\prime}+2\pi)$
is non-trivial. If we define the following
complexified coordinate $w=\eta+ix$, then the identification is translated
into $(w,z)\sim(we{}^{2\pi(R_{+}+iR_{-})},ze^{2\pi R_{+}})$. From
this perspective, we can take the fundamental region as $1\leq\left|w\right|^{2}+z^{2}\leq e^{2\pi R_{+}}$
and the horizon $r=R_{+}$ is mapped to $w=0$ and $1\leq z\leq e{}^{2\pi R_{+}}$.

Let us insert the ETW branes in this setup. 
The position of ETW
branes with a tension $T$ is determined by the Neumann boundary condition in the AdS/BCFT:
\begin{equation}
K_{ab}-Kh_{ab}=-Th_{ab}.  \label{neuads}
\end{equation}

In the Poincare coordinate, the solution to the boundary condition (\ref{neuads}) is solved as follows \citep{Akal:2020wfl}:
\begin{equation}
(z-\alpha)^{2}+(x-p)^{2}+(\eta-q)^{2}=\beta^{2},\label{2.6}
\end{equation}
where the tension is given by $T=\frac{\alpha}{\beta}$. Here ETW
branes, where $\phi^{\prime}$ is constant, are just a sphere of radius
1 for $\phi^{\prime}=0$ and $e^{\pi R_{+}}$ for $\phi^{\prime}=\pi$.
Next, we calculate the action
\begin{eqnarray}
\thickmuskip=0mu
\medmuskip=0mu
\thinmuskip=0muS&=&-\frac{1}{16\pi G}\int_{N}\sqrt{g}(R+2)d^{3}x-\frac{1}{8\pi G}\int_{Q}\sqrt{h}(K-T)d^{2}x \nonumber \\
&-&\frac{1}{8\pi G}\int_{M}\sqrt{h}Kd^{2}x+\frac{k}{8\pi G}S_{ct},
\end{eqnarray}
where $N$ is the three-dimensional bulk region, $Q$ is the ETW brane
and $M$ is the conformal boundary placed at $r=R$, which we later take
$R\rightarrow\infty$. $S_{ct}=\int_{M}\sqrt{h}$ is a counter-term
constructed only from induced geometric quantities \citep{Balasubramanian:1999re}.

Consider computing the induced metric and the extrinsic
curvature at $r=R$ surface. After some algebras, we get
\begin{equation}
h_{ab}=\left(\begin{array}{ccc}
A' & 0 & B'\\
0 & 0 & 0\\
B & 0 & C'
\end{array}\right),
\end{equation}
where
\begin{eqnarray}
A'&=&=\frac{(r^{2}-R_{+}^{2})(r^{2}+R_{-}^{2})}{R^{2}}+R^{2}\left(\frac{R_{-}}{R_{+}}-\frac{R_{+}R_{-}}{R^{2}}\right)^{2}, \nonumber\\
B'&=&\frac{R_{-}}{R_{+}}R^{2}-R_{+}R_{-},\nonumber\\
C'&=&R^{2},
\end{eqnarray}
and the extrinsic curvature reads
\begin{equation}
K_{ab}=\left(\begin{array}{cc}
\frac{R_{+}^{2}+R_{-}^{2}}{R_{+}^{2}} & \frac{R_{-}}{R_{+}}\\
\frac{R_{-}}{R_{+}} & 1
\end{array}\right)\sqrt{\left(R^{2}-R_{+}^{2}\right)\left(r^{2}+R_{-}^{2}\right)}.
\end{equation}
From this we find
\begin{equation}
\begin{array}{cc}
\det(h_{ab})=(R^{2}-R_{+}^{2})(R^{2}+R_{-}^{2}),\\
K=\frac{2R^{2}-R_{+}^{2}+R_{-}^{2}}{\sqrt{(R^{2}-R_{+}^{2})(R^{2}+R_{-}^{2})}}.
\end{array}
\end{equation}
Now, the Einstein-Hilbert action is just the volume integral:
\begin{equation}
\thickmuskip=0mu
\medmuskip=0mu
\thinmuskip=0mu-\frac{1}{16\pi G}\int_{R_{+}}^{R}dr\int_{0}^{\beta}dt\int_{0}^{\pi}d\phi^{\prime}(-4r)=\frac{4\pi\beta}{16\pi G}\left(\frac{R^{2}-R_{+}^{2}}{2}\right).
\end{equation}
We can also evaluate the GHY term at $r=R$:
\begin{eqnarray}
\thickmuskip=0mu
\medmuskip=0mu
\thinmuskip=0mu
&&-\frac{1}{8\pi G}\int_{0}^{\beta}\!dt\int_{\frac{2\pi}{\beta}\left(1-\frac{t}{\beta}\right)}^{\frac{2\pi}{\beta}\left(1-\frac{t}{\beta}\right)+\pi}\!d\phi^{\prime}\left[\sqrt{(R^{2}-R_{+}^{2})(R^{2}+R_{-}^{2})}\right.\nonumber\\
&&\left.\left(\frac{2R^{2}-R_{+}^{2}+R_{-}^{2}}{(R^{2}-R_{+}^{2})(R^{2}+R_{-}^{2})}\!\right)\!\right]\!=\!-\frac{\beta}{8G}(2R^{2}-R_{+}^{2}+R_{-}^{2}).\nonumber\\
\end{eqnarray}

Then, we determine the counterterm. That should be constructed from
a geometric quantity of the boundary surface. Here as the most simple one
($k=1)$ we take
\begin{equation}
S_{ct}=\int_{M}d^{2}x\sqrt{h}=\beta\pi\sqrt{(R^{2}-R_{+}^{2})(R^{2}+R_{-}^{2})}.
\end{equation}
We combine the above results altogether and we get:
\begin{equation}
S=-\frac{\beta}{8G}\left(R^{2}+R_{-}^{2}-\sqrt{(R^{2}-R_{+}^{2})(R^{2}+R_{-}^{2})}\right).
\end{equation}
If we take $R\rightarrow\infty$ limit this becomes
\begin{equation}
S\simeq-\frac{\beta\left(R_{+}^{2}+R_{-}^{2}\right)}{16G}=-\frac{\pi R_{+}}{8G}=-\frac{\pi^{2}\beta}{4G(\beta^{2}+\theta^{2})}.
\end{equation}
This is what we expected because we chose the position of the ETW branes so
that the volume of the bulk space becomes a half of the original volume.
In the full AdS case, this is done in \citep{Maldacena:1998bw}, which
is just twice of the action in our calculation.

\subsection{BTZ with ETW branes for general values of tension $T$}

Next, we consider the general case with non-vanishing $T$. It is natural to assume the rotational symmetry i.e. setting $p=q=0$ in (\ref{2.6}). Therefore, we will
consider the following equation
\begin{equation}
\thickmuskip=0mu
\medmuskip=0mu
\thinmuskip=0mu\left(\!\left(\!\frac{R_{+}^{2}+R_{-}^{2}}{r^{2}+R_{-}^{2}}\right)^{\frac{1}{2}}e^{R_{+}\phi^{\prime}}-\alpha\right)^{2}\!+\!\left(\!\frac{r^{2}-R_{+}^{2}}{r^{2}+R_{-}^{2}}\!\right)e^{2R_{+}\phi^{\prime}}=\beta^{2}.
\end{equation}

Since the brane is anchored at $\phi^{\prime}=0$ and $\pi$, if we take $r\rightarrow\infty$,
then the brane equation becomes
\begin{eqnarray}
\alpha^{2}+1=\beta^{2} & \ \  (\phi^{\prime}=0),\nonumber\\
\alpha^{2}+\exp(2\pi R_{+})=\beta^{2} & \ \ (\phi^{\prime}=\pi).
\end{eqnarray}
We note that $T$ has range $-1<T<1$ from the above constraint. Using
this we can determine the brane configuration: in the case where
the brane is anchored at $\phi^{\prime}=0$, the equation becomes
\begin{eqnarray}
\thickmuskip=0mu
\medmuskip=0mu
\thinmuskip=0mu
\phi^{\prime}&=&\frac{1}{R_{+}}\log\left(\frac{T}{\sqrt{1-T^{2}}}\sqrt{\frac{R_{+}^{2}+R_{-}^{2}}{r^{2}+R_{-}^{2}}}\right.\nonumber\\
&+&\left.\sqrt{\frac{T^{2}}{1-T^{2}}(\frac{R_{+}^{2}+R_{-}^{2}}{r^{2}+R_{-}^{2}})+1}\right)
\end{eqnarray}
and in the case where the brane is anchored at $\phi^{\prime}=\pi$,
we have
\begin{eqnarray}
\thickmuskip=0mu
\medmuskip=0mu
\thinmuskip=0mu
\phi^{\prime}&=&\frac{1}{R_{+}}\log\left\{e^{\pi R_{+}}\left(\frac{T}{\sqrt{1-T^{2}}}\sqrt{\frac{R_{+}^{2}+R_{-}^{2}}{r^{2}+R_{-}^{2}}}\right.\right. \nonumber\\
&+&\left.\left.\sqrt{\frac{T^{2}}{1-T^{2}}(\frac{R_{+}^{2}+R_{-}^{2}}{r^{2}+R_{-}^{2}})+1}\right)\right\}.
\end{eqnarray}
Then the Einstein-Hilbert
action can be calculated directly:
\begin{eqnarray}
\thickmuskip=0mu
\medmuskip=0mu
\thinmuskip=0mu 
S_{EH}&=&-\frac{1}{16\pi G}\int_{R_{+}}^{R_{reg}}dr\int_{0}^{\beta}dt\int d\phi^{\prime}(-4r)\nonumber\\
&=&\frac{\beta}{8G}(R_{reg}^{2}-R_{+}^{2}).
\end{eqnarray}
This is the same result as $T=0$ case.

Next, we evaluate the brane action. In
three dimensions, the extrinsic curvature satisfies $K=2T$.
For the brane anchored at $\phi^{\prime}=0$, the induced metric reads:
\begin{eqnarray}
\thickmuskip=0mu
\medmuskip=0mu
\thinmuskip=0mu 
h_{rr}&=&\frac{r^{4}T^{2}(R_{+}^{2}+R_{-}^{2})}{R_{+}^{2}(r^{2}+R_{-}^{2})^{2}(T^{2}R_{+}^{2}+R_{-}^{2}-r^{2}(T^{2}-1))}\nonumber\\
&+&\frac{r^{2}}{\left(r^{2}-R_{+}^{2}\right)\left(r^{2}+R_{-}^{2}\right)},\\
h_{tr}\!&=&\!\!\frac{R_{-}Tr(R_{+}^{2}-r^{2})\sqrt{R_{+}^{2}+R_{-}^{2}}}{\!R_{+}^{2}\!(r^{2}+R_{-}^{2})\!\sqrt{\!r^{2}(1-T^{2})\!+R_{+}^{2}T^{2}+R_{-}^{2}}},\\
h_{tt}&=&\!\frac{\left(r^{2}-R_{+}^{2}\right)\left(r^{2}+R_{-}^{2}\right)}{r^{2}}\!+\!\left(\frac{R_{-}}{R_{+}}-\frac{R_{+}R_{-}}{r^{2}}\right)^{2}\!r^{2}.\nonumber\\
\end{eqnarray} \\
From this we find
\begin{equation}
\det(h_{ab})=\frac{r^{2}(R_{+}^{2}+R_{-}^{2})}{R_{+}^{2}(r^{2}(1-T^{2})+R_{+}^{2}T^{2}+R_{-}^{2})}.
\end{equation}
On the other hand, in the case where the brane is anchored at $\phi^{\prime}=\pi$ ,
we can repeat the same thing and the result is the same as above.
The brane action can be written as
\begin{eqnarray}
\thickmuskip=0mu
\medmuskip=0mu
\thinmuskip=0mu 
&&S_{brane}\nonumber\\
&=&\frac{T}{8\pi G}\int_{R_{+}}^{R_{reg}}dr\int_{0}^{\beta}dt\sqrt{\frac{r^{2}(R_{+}^{2}+R_{-}^{2})}{R_{+}^{2}(r^{2}(1-T^{2})+R_{+}^{2}T^{2}+R_{-}^{2})}}.\nonumber\\
\end{eqnarray}

However, the brane contribution for $\phi^{\prime}=0$ and $\phi^{\prime}=\pi$ is canceled as we note in the calculation of the Einstein-Hilbert action; the
brane is curved in the opposite direction with respect to the bulk
space. For the other terms on the conformal boundary, we have just
the same term as in $T=0$ case because in $r\rightarrow\infty$ limit
$\phi^{\prime}$ goes to $0$ and $\pi$.

Combining them, we obtain
\begin{eqnarray}
\thickmuskip=0mu
\medmuskip=0mu
\thinmuskip=0mu 
S&=&\frac{\beta}{8G}(R_{reg}^{2}-R_{+}^{2})\nonumber\\
&-&\frac{\beta}{8G}(2R_{reg}^{2}-R_{+}^{2}+R_{-}^{2})\nonumber\\
&+&\frac{k\beta}{8G}\sqrt{(R_{reg}^{2}-R_{+}^{2})(R_{reg}^{2}+R_{-}^{2})}\nonumber\\
&\rightarrow&-\frac{\beta\left(R_{+}^{2}+R_{-}^{2}\right)}{16G}, 
\end{eqnarray}
where we take $R_{reg}\rightarrow\infty$.

If we assume the tensions of the two branes take different values: $T_{0}$ and $T_{\pi}$, where the branes are
anchored at $\phi^{\prime}=0,\pi$. For the Einstein-Hilbert action
part the integral becomes

\begin{eqnarray}
\thickmuskip=0mu
\medmuskip=0mu
\thinmuskip=0mu 
S_{EH}&=&-\frac{1}{16 G}\int_{R_{+}}^{R_{reg}}dr\int_{0}^{\beta}dt(-4r)\nonumber\\
&& \cdot\log\!\left[\!\frac{ \!\left(\!\frac{T_{\pi}}{\sqrt{1-T_{\pi}^{2}}}\!\sqrt{\frac{R_{+}^{2}+R_{-}^{2}}{r^{2}+R_{-}^{2}}}\!+\!\sqrt{\frac{T_{\pi}^{2}}{1-T_{\pi}^{2}}(\frac{R_{+}^{2}+R_{-}^{2}}{r^{2}+R_{-}^{2}})\!+\!1}\right)}{\left(\!\frac{T_{0}}{\sqrt{1-T_{0}^{2}}}\!\sqrt{\frac{R_{+}^{2}+R_{-}^{2}}{r^{2}+R_{-}^{2}}}\!+\!\sqrt{\frac{T_{0}^{2}}{1-T_{0}^{2}}(\frac{R_{+}^{2}+R_{-}^{2}}{r^{2}+R_{-}^{2}})\!+\!1}\right)\!}\!\right].\nonumber\\
\end{eqnarray}

After performing the integral, we finally obtain
\begin{eqnarray}
\thickmuskip=0mu
\medmuskip=0mu
\thinmuskip=0mu 
S&=&-\frac{\beta\left(R_{+}^{2}+R_{-}^{2}\right)}{16G}\nonumber\\
&-&\frac{\beta\left(R_{+}^{2}+R_{-}^{2}\right)}{16\pi GR_{+}}\!\left(\log\left(\frac{1+T_{\pi}}{1-T_{\pi}}\right)\!-\!\log\left(\frac{1+T_{0}}{1-T_{0}}\!\right)\!\right),\nonumber\\
\end{eqnarray}
where we omit the divergent term, which is proportional to $R_{reg}$.
For the non-rotating BTZ, this result matches with that in \citep{Fujita:2011fp}.

\section{A One-loop partition function in the thermal AdS with a tensionless ETW
brane}

In this section, we will use the explicit form of the heat kernel
presented in \citep{Giombi:2008vd}, though our presentation will be brief. Please refer to the original paper if necessary.

\subsection{A review of the heat kernel method}

The heat kernel method is a convenient way of calculating a one-loop
partition function. We follow the analysis in \citep{David:2009xg,Giombi:2008vd,Gopakumar:2011qs,Mann:1996ze,Vassilevich:2003xt}.
Let us consider calculating the partition function of a free scalar $\phi$:
\begin{equation}
Z=\int D\phi\,e^{-S(\phi)}.
\end{equation}

The action of the scalar field can be rewritten as
\begin{equation}
S(\phi)=\int_{M}d^{3}x\,\sqrt{g}\phi\Delta\phi,
\end{equation}
where we omit indices of tensorial structure. Since we are considering a compact space, $\Delta$ has a discrete set of eigenvalues $\lambda_{n}$.
The one-loop partition function is expressed as follows:
\begin{equation}
S^{(1)}=-\frac{1}{2}\log\det(\Delta)=-\frac{1}{2}\underset{n}{\sum}\log\lambda_{n}.
\end{equation}
If we consider a non-compact space, the spectrum becomes continuous
and the one-loop partition function is divergent, which is proportional
to the volume. This divergence can be absorbed by the renormalization
of the Newton constant.

The heat kernel is defined as

\begin{equation}
K(t,x,y)=\underset{n}{\sum}e^{-\lambda_{n}t}\psi_{n}(x)\psi_{n}(y),
\end{equation}
which we usually call a propagator. We can normalize the eigenfunctions
as

\begin{equation}
\begin{array}{cc}
\underset{n}{\sum}\psi_{n}(x)\psi_{n}(y)=\delta^{3}(x-y),\\
\int_{M}d^{3}x\underset{n}{\sum}\sqrt{g}\psi_{n}(x)\psi_{m}(x)=\delta_{nm}.
\end{array}
\end{equation}
The trace of the heat kernel is given by

\begin{equation}
\int_{M}d^{3}x\underset{n}{\sum}\sqrt{g}K(t,x,x)=\underset{n}{\sum}e^{-\lambda_{n}t}.
\end{equation}
Using this we can compute the 1-loop partition function as an integral
over $t$:

\begin{equation}
S^{(1)}=-\frac{1}{2}\underset{n}{\sum}\log\lambda_{n}=\frac{1}{2}\int_{+0}^{\infty}\frac{dt}{t}\int_{M}d^{3}x\underset{n}{\sum}\sqrt{g}K(t,x,x).
\end{equation}

We can show the above equation by differentiating with respect to
$\lambda_{n}$. Note that it is an identity up to an infinite constant.
The point is that $K$ satisfies the heat conduction equation

\begin{equation}
\left(\partial_{t}+\Delta_{x}\right)K(t,x,y)=0,
\end{equation}
with a boundary condition at $t=0$

\begin{equation}
K(0,x,y)=\delta(x,y).
\end{equation}

\subsection{A One-loop partition function in thermal AdS with ETW brane}

In this section, we apply the calculation in \citep{Giombi:2008vd}
to our ETW brane setup. Now consider Poincare $AdS_{3}$ whose metric
is given by

\begin{equation}
ds^{2}=\frac{dy^{2}+dzd\overline{z}}{y^{2}}.
\end{equation}
Here an ETW brane is placed at $\Re(z)=0$ and the bulk region is defined
as $\Re(z)>0$. We note that the ETW brane is connected in the bulk.
Since the AdS space is maximally isometric, the geodesic distance
$r(x,x^{\prime})$ depends only on the chordal distance $u(x,x^{\prime})$:

\begin{equation}
\begin{array}{cc}
r(x,x^{\prime})=\mbox{arccosh}\left(1+u(x,x^{\prime})\right),\\
\\
\end{array}
\end{equation}
where

\begin{equation}
u(x,x^{\prime})=\frac{\left(y-y^{\prime}\right)^{2}+\left|z-z^{\prime}\right|^{2}}{2yy^{\prime}}.
\end{equation}

A thermal AdS can be obtained from an AdS space using the following identification

\begin{equation}
\left(y,z\right)\sim\left(\left|q\right|^{-1}y,q^{-1}z\right),
\end{equation}
where $q=e^{2\pi i\tau}$ and $\tau=\tau_{1}+i\tau_{2}$. In the non-zero
$\tau_{1}$ case, the boundary of the BCFT wraps around the torus
for many times and in this case the region on which BCFT lives, is
not clear because the region is not surrounded by the boundaries. To
cure this problem later we will restrict to the special case $q=\overline{q}$. Now we consider applying the method of images to the heat kernel method.
The tensionless ETW brane is inserted at $\Re(z)=0$. Then, if we consider
a mirror position, $z$ and $\overline{z}$ are mapped to $-\overline{z}$
and $-z$, respectively. Then, one-loop partition function becomes
\begin{eqnarray}
\thickmuskip=0mu
\medmuskip=0mu
\thinmuskip=0mu 
S^{(1)}&=&\frac{1}{2}\int_{+0}^{\infty}\frac{dt}{t}\int_{thermal\,AdS}d^{3}x\underset{n}{\sum}\sqrt{g}\left(K^{\mathbb{\mathbb{H}}/\mathbb{Z}}(t,x,x)\right.\nonumber\\
&+&\left.K^{\mathbb{\mathbb{H}}/\mathbb{Z}}(t,x^{mirror},x)\right).\label{eq:-3}
\end{eqnarray}
In the tensionless case the boundary condition for the metric is given
by $K_{ab}=0$. We note that we treat this condition as an off-shell boundary condition. This means that we must impose the following boundary condition for
the perturbation of the metric:
\begin{equation}
\partial_{x}h_{ij}=0.
\end{equation}

Since the heat kernel on a thermal AdS can be obtained using the method of
images from that of an AdS space, we get
\begin{eqnarray}
\thickmuskip=0mu
\medmuskip=0mu
\thinmuskip=0mu 
S^{(1)}&=&\frac{1}{2}\int_{+0}^{\infty}\frac{dt}{t}\int_{thermal\,AdS}d^{3}x\underset{n}{\sum}\sqrt{g}\left(K^{\mathbb{\mathbb{H}}}(t,x,\gamma^{n}x)\right.\nonumber\\
&+&\left.K^{\mathbb{\mathbb{H}}}(t,x^{mirror},\gamma^{n}x)\right).
\end{eqnarray}
For a later convenience we will use a different coordinate:
\begin{equation}
y=\rho\sin\theta, \ \ \  z=\rho\cos\theta e^{i\phi},
\end{equation}
where $1\leq\rho\leq e^{2\pi\tau_{2}}$ , $0\leq\theta\leq\frac{\pi}{2}$
and $-\frac{\pi}{2}\leq\phi\leq\frac{\pi}{2}$ . In terms of this
coordinate the geodesic distances are given by
\begin{equation}
\begin{array}{cc}
r(x,\gamma^{n}x)=\mbox{arccosh}\left(\frac{\cosh\,\beta}{\sin^{2}\theta}-\frac{\cos\,\alpha}{\tan^{2}\theta}\right),\\
r(x^{mirror},\gamma^{n}x)=\mbox{arccosh}\left(\frac{\cosh\,\beta}{\sin^{2}\theta}+\frac{\cos(2\phi-\alpha)}{\tan^{2}\theta}\right),
\end{array}
\end{equation}
where we defined $\alpha=2\pi n\tau_{1}$ and $\beta=2\pi n\tau_{2}$.

\subsection{A One-loop partition function of a scalar field}

The heat kernel of a scalar field on an $AdS_{3}$ space is given by

\begin{equation}
K^{H_{3}}(t,r(x,x^{\prime}))=\frac{e^{-(m^{2}+1)t-\frac{r^{2}}{4t}}}{(4\pi t)^{\frac{3}{2}}}\frac{r}{\sinh(r)}.
\end{equation}
Let us consider calculating the ordinary (not the mirror) part. A one-loop
determinant can be recast as an integral
\begin{equation}
\begin{array}{cc}
-\log\det\Delta=Vol(H_{3}/\mathbb{Z})\int_{0}^{\infty}\frac{dt}{t}\frac{e^{-(m^{2}+1)t}}{(4\pi t)^{\frac{3}{2}}}\\
+\underset{n\neq0}{\sum}\int_{0}^{\infty}\frac{dt}{t}\int\frac{d\rho d\theta d\phi\cos\theta}{\rho\sin^{3}\theta}\frac{e^{-(m^{2}+1)t-\frac{r^{2}}{4t}}}{(4\pi t)^{\frac{3}{2}}}\frac{r}{\sinh(r)},
\end{array}
\end{equation}
where we split into the zero mode part and non-zero mode $n\neq 0$. The first term can be easily regularized:
\begin{equation}\thickmuskip=0mu
\medmuskip=0mu
\thinmuskip=0mu 
\int_{0}^{\infty}\frac{dt}{t}\frac{e^{-(m^{2}+1)t}}{(4\pi t)^{\frac{3}{2}}}=\frac{\left(m^{2}+1\right)^{\frac{3}{2}}}{8\pi^{\frac{3}{2}}}\int_{0}^{\infty}dkk^{-\frac{5}{2}}e^{-k}=\frac{\left(m^{2}+1\right)^{\frac{3}{2}}}{6\pi}.
\end{equation}
The second term can be calculated directly. Firstly, we change the
variable from $\theta$ to $r$ and we get

\begin{equation}
\begin{array}{cc}
\underset{n\neq0}{\sum}\int_{0}^{\infty}\frac{dt}{t}\int\frac{d\rho d\theta d\phi\cos\theta}{\rho\sin^{3}\theta}\frac{e^{-(m^{2}+1)t-\frac{r^{2}}{4t}}}{(4\pi t)^{\frac{3}{2}}}\frac{r}{\sinh(r)}\\
=\underset{n\neq0}{\sum}\int_{0}^{\infty}\frac{dt}{t}\int\frac{d\rho drd\phi}{\rho}\frac{e^{-(m^{2}+1)t-\frac{r^{2}}{4t}}}{(4\pi t)^{\frac{3}{2}}}\frac{r}{2\left(\cosh\beta-\cos\alpha\right)}.
\end{array}
\end{equation}
Integrating over $r$, $\phi$ and $t$ in order, we can reach
the final answer:

\begin{equation}
\begin{array}{cc}
\underset{n\neq0}{\sum}\int_{0}^{\infty}\frac{dt}{t}\int\frac{d\rho drd\phi}{\rho}\frac{e^{-(m^{2}+1)t-\frac{r^{2}}{4t}}}{(4\pi t)^{\frac{3}{2}}}\frac{r}{2\left(\cosh\beta-\cos\alpha\right)}\\
=\underset{n\neq0}{\sum}\frac{\sqrt{\pi}\tau_{2}}{4\left(\cosh\beta-\cos\alpha\right)}\int_{0}^{\infty}\frac{dt}{t}\frac{e^{-(m^{2}+1)t-\frac{\left(2\pi n\tau_{2}\right)^{2}}{4t}}}{t^{\frac{3}{2}}}\\
=\underset{n\neq0}{\sum}\frac{e^{-2\pi n\tau_{2}\sqrt{m^{2}+1}}}{4n\left(\cosh\beta-\cos\alpha\right)}\\
=\sum_{n=1}^\infty\frac{\left|q\right|^{n\left(1+\sqrt{m^{2}+1}\right)}}{n\left|1-q^{n}\right|^{2}}.
\end{array}
\end{equation}

Now let us consider the mirror part. Firstly, we consider the $n=0$ case.
The strategy is that we firstly fix $\phi$ and integrate $\theta$
or $r$ and then integrate over other variables. This leads to the
divergent result even though we use a regularization by a gamma function:

\begin{equation}
\begin{array}{cc}
\int_{0}^{\infty}\frac{dt}{t}\int\frac{d\rho d\theta d\phi\cos\theta}{\rho\sin^{3}\theta}\frac{e^{-(m^{2}+1)t-\frac{r^{2}}{4t}}}{(4\pi t)^{\frac{3}{2}}}\frac{r}{\sinh(r)}\\
=-\frac{\tau_{2}\sqrt{m^{2}+1}}{2}\int_{-\frac{\pi}{2}}^{\frac{\pi}{2}}\frac{d\phi}{1+\cos(2\phi)}.
\end{array}
\end{equation}
However, this divergence can be eliminated by the renormalization
of Newton constant, so we will ignore this term.

Next consider the $n\neq0$ case. The trick is almost the same, but the
situation is slightly changed. The integral over $\phi$ yields
\begin{equation}
\begin{array}{cc}
\underset{n\neq0}{\sum}\int_{0}^{\infty}dt2\pi\tau_{2}\frac{e^{-(m^{2}+1)t-\frac{\left(2\pi n\tau_{2}\right)^{2}}{4t}}}{(4\pi t)^{\frac{3}{2}}}\int_{-\frac{\pi}{2}}^{\frac{\pi}{2}}\frac{d\phi}{\left|\cos\left(2\phi-\alpha\right)+\cosh\left(\beta\right)\right|}\\
=\underset{n\neq0}{\sum}\int_{0}^{\infty}dt\,2\pi\tau_{2}\frac{e^{-(m^{2}+1)t-\frac{\left(2\pi n\tau_{2}\right)^{2}}{4t}}}{(4\pi t)^{\frac{3}{2}}}\frac{\pi}{\sinh(\beta)}\\
=\sum_{n=1}^{\infty}\frac{\left|q\right|^{n\left(1+\sqrt{m^{2}+1}\right)}}{n\left(1-\left|q\right|^{2n}\right)}.
\end{array}
\end{equation}

\subsection{A one-loop partition function for a vector field}

In this section we consider the vector field contribution to the partition
function. For a later convenience we define a set of new real coordinates:

\begin{equation}
\begin{array}{cc}
x=\Re(z)=\rho\cos\theta\cos\phi,\\
\eta=\Im(z)=\rho\cos\theta\sin\phi.
\end{array}
\end{equation}
The propagator has two indices and is invariant under an exchange of those
indices. Thus we can expand the kernel by the basis of (1,1) symmetric
bi-tensor \citep{DHoker:1999bve}:

\begin{equation}
K_{\mu\nu^{\prime}}(t,x,x^{\prime})=F(t,u)\partial_{\mu}\partial_{\nu^{\prime}}u+\partial_{\mu}\partial_{\nu^{\prime}}S(t,u),
\end{equation}
where 
\begin{equation}
\begin{array}{cc}
F(t,r)=-\frac{e^{-\frac{r^{2}}{4t}}}{(4\pi t)^{\frac{3}{2}}}\frac{r}{\sinh r},\\
S(t,r)=\frac{4}{(4\pi)^{\frac{3}{2}}}\frac{e^{-\frac{r^{2}}{4t}}}{\sinh r}\sqrt{t}\int_{0}^{1}d\xi e^{-t(1-\xi)^{2}}\sinh(r\xi).
\end{array}
\end{equation}
The next step is to calculate the derivatives of $u$. In the Poincare
coordinate it can be explicitly given by
\begin{eqnarray}
&&\partial_{\mu}\partial_{\nu^{\prime}}u\nonumber\\
&=&-\frac{1}{z_{0}w_{0}}\left\{ \delta_{\mu\nu^{\prime}}\!+\!\frac{(z-w)_{\mu}}{w_{0}}\delta_{\nu^{\prime}0}+\frac{(w-z)_{\nu^{\prime}}}{z_{0}}\delta_{\mu0}\!-\!u\delta_{\mu0}\delta_{\nu^{\prime}0}\right\}.\nonumber\\
\end{eqnarray}
Here we note that $z=x$ or $z=x^{mirror}$, and $w=\gamma^{n}x$ . We
will present a detailed calculation of these derivatives in the appendix A.
Let us consider the ordinary (not the mirror image) contribution. 
Using those results, we can compute the kernel:
\begin{eqnarray}
\thickmuskip=0mu
\medmuskip=0mu
\thinmuskip=0mu 
&&\int_{0}^{\infty}\frac{dt}{t}\int d^{3}x\sqrt{g}\sum_{n}g^{\mu\rho}\frac{\partial(\gamma^{n}x)^{\nu^{\prime}}}{\partial x^{\rho}}K_{\mu\nu^{\prime}}^{\mathbb{\mathbb{H}}}(t,r(x,\gamma^{n}x))\nonumber\\
&=&\int_{0}^{\infty}\frac{dt}{t}\int   d^{3}x\sqrt{g}\underset{n}{\sum}
\left\{
\frac{d^{2}}S{du^{2}}\left\{ (e^{\beta}-\cosh r)(e^{-\beta}\cosh r)\right.
\right.\nonumber\\
&-&\left.\left.2\cos\alpha(\cosh r-\cosh\beta)\right\}\right. \nonumber\\
&+&\left.(F+(\frac{dS}{du}))\left(\cosh r-2\cosh\beta-2\cos\alpha\right)\right\}\nonumber\\
&=&Vol(\mathbb{H}/\mathbb{Z})\int_{0}^{\infty}\frac{dt}{t}\frac{(e^{-t}+2+4t)}{(4\pi t)^{\frac{3}{2}}}\nonumber\\
&+&\sum_{n=0}^{\infty}\int_{0}^{\infty}\frac{dt}{t}\frac{2\pi^{2}\tau_{2}}{(\cosh\beta-\cos\alpha)}\frac{e^{-\frac{\beta^{2}}{4t}}}{4\pi^{\frac{3}{2}}\sqrt{t}}(2\cos\alpha+e^{-t}).\nonumber\\
\end{eqnarray}
The first term can be regularized by considering a massless limit
of a massive field:
\begin{eqnarray}
\thickmuskip=0mu
\medmuskip=0mu
\thinmuskip=0mu 
&&Vol(\mathbb{H}/\mathbb{Z})\int_{0}^{\infty}\frac{dt}{t}\frac{(e^{-t}+2+4t)}{(4\pi t)^{\frac{3}{2}}}\nonumber\\
&=&\underset{m\rightarrow0}{lim}Vol(\mathbb{H}/\mathbb{Z})\int_{0}^{\infty}\frac{dt}{t}\frac{e^{-m^{2}t}(e^{-t}+2+4t)}{(4\pi t)^{\frac{3}{2}}}\nonumber\\
&=&Vol(\mathbb{H}/\mathbb{Z})\frac{1}{(4\pi)^{\frac{3}{2}}}(\frac{4\sqrt{\pi}}{3}-8\sqrt{\pi})=-\frac{5}{6\pi}Vol(\mathbb{H}/\mathbb{Z}).\nonumber\\
\end{eqnarray}
The second term can be calculated straightforwardly:
\begin{equation}
\begin{array}{cc}
\sum_{n=1}^{\infty}\int_{0}^{\infty}\frac{dt}{t}\frac{2\pi^{2}\tau_{2}}{(\cosh\beta-\cos\alpha)}\frac{e^{-\frac{\beta^{2}}{4t}}}{4\pi^{\frac{3}{2}}\sqrt{t}}(2\cos\alpha+e^{-t})\\
=\sum_{n=1}^{\infty}\frac{2\cos\alpha+e^{-\beta}}{2n(\cosh\beta-\cos\alpha)}\\
=\sum_{n=1}^{\infty}\frac{q^{n}+\overline{q}^{n}+\left|q\right|^{2n}}{n\left|1-q^{n}\right|^{2}}.
\end{array}
\end{equation}
If we omit the term, which is proportional to the volume, the answer
is just a half of the original result as we expected.

Next, we consider the mirror image part. Since the brane is inserted at $x=0$,
we have $x_{mirror}=-x$, $\eta_{mirror}=\eta$ and y$_{mirror}=y$.
The derivatives of $u$ are also slightly modified and we present
the results in the appendix A. Computations analogous to the previous ones lead to

\begin{eqnarray}
\thickmuskip=0mu
\medmuskip=0mu
\thinmuskip=0mu 
&&\int_{0}^{\infty}\frac{dt}{t}\int d^{3}x\sqrt{g}\underset{n}{\sum}g^{\mu\rho}\frac{\partial(\gamma^{n}x)^{\nu^{\prime}}}{\partial x^{\rho}}K_{\mu\nu^{\prime}}^{\mathbb{\mathbb{H}}}(t,r(x_{mirror},\gamma^{n}x))\nonumber\\
&=&\int_{0}^{\infty}\!\frac{dt}{t}\!\int\! d^{3}x\!\sqrt{g}\!\sum_{n}\left[ \left(\frac{d^{2}S}{du^{2}}\right)
\left\{(e^{\beta}-\cosh r)(e^{-\beta}-\cosh r)\right. \right.\nonumber\\
&-&\left.\left.2\cos\alpha(\cosh r-\cosh\beta)\right\}\right. \nonumber\\
&+&\left.(F+(\frac{dS}{du}))(\cosh r-2\cosh\beta-2\cos\alpha)\right].
\end{eqnarray}

The above form is the same as the previous result, except that we have different geodesic distances:
\begin{equation}
r=arccosh\left(\frac{\cosh\beta}{\sin^{2}\theta}+\frac{\cos(2\phi-\alpha)}{\tan^{2}\theta}\right).
\end{equation}
This distance $r$ depends not only $\theta$ , but also $\phi$ .
To proceed, we fix $\phi$ and integrate $\theta$ (or $r$) firstly.
The integral over $r$ gives the same as the original calculation
because it is independent of $\phi$ :
\begin{eqnarray}
\thickmuskip=0mu
\medmuskip=0mu
\thinmuskip=0mu 
&&\int_{0}^{\infty}\frac{dt}{t}\int d^{3}x\sqrt{g}\sum_{n}\Biggl[\left(\frac{d^{2}S}{du^{2}}\right)\nonumber\\
&& \left\{(e^{\beta}-\cosh r)(e^{-\beta}-\cosh r)-2\cos\alpha(\cosh r-\cosh\beta)\right\} \nonumber\\
&+&\left.(F+(\frac{dS}{du}))(\cosh r-2\cosh\beta-2\cos\alpha)\right]\nonumber\\
&=&\int_{0}^{\infty}\!\frac{dt}{t}\!\sum_{n}\!2\pi\tau_{2}\!\int_{-\frac{\pi}{2}}^{\frac{\pi}{2}}\!d\phi\!\left[\frac{1}{2(\cosh\beta+\cos(2\phi-\alpha))}\right.\nonumber\\
&&\frac{e^{-\frac{\beta^{2}}{4t}}}{4\pi^{\frac{3}{2}}\sqrt{t}}(2\cos\alpha+e^{-t})\Biggr].
\end{eqnarray}

The integral over $\phi$ can also be evaluated:
\begin{equation}
\begin{array}{cc}
\int_{-\frac{\pi}{2}}^{\frac{\pi}{2}}d\phi\frac{1}{(\cosh\beta+\cos(2\phi-\alpha))}
=\frac{\pi}{\sinh\beta}.
\end{array}
\end{equation}

The mirror image contribution finally becomes
\begin{equation}
\begin{array}{cc}
\thickmuskip=0mu
\medmuskip=0mu
\thinmuskip=0mu \int_{0}^{\infty}\frac{dt}{t}\int d^{3}x\sqrt{g}\underset{n}{\sum}g^{\mu\rho}\frac{\partial(\gamma^{n}x)^{\nu^{\prime}}}{\partial x^{\rho}}K_{\mu\nu^{\prime}}^{\mathbb{\mathbb{H}}}(t,r(x_{mirror},\gamma^{n}x))\\
=\sum_{n=1}^{\infty}\frac{2\cos\alpha+e^{-\beta}}{2n\sinh\beta}\\
=\sum_{n=1}^{\infty}\frac{q^{n}+\overline{q}^{n}+\left|q\right|^{2n}}{n(1-\left|q\right|^{2n})}.
\end{array}
\end{equation}

\subsection{A one-loop partition function of a symmetric spin-2 field}

In the calculation of the vector field one-loop partition function 
we see that the trace of the kernel in the mirror part is the same as that in the ordinary part. This seems to be somewhat
miraculous at a first sight.  However this is not surprising because the trace of
the kernel can only depend on geodesic distance $r$. 

The kernel of the symmetric spin-2 field can be expanded by the basis of (2,2) symmetric tensors \citep{DHoker:1999bve,Giombi:2008vd}.
Since they are complicated, we do not want to write explicitly here.

The one-loop determinant of a symmetric traceless tensor is given by
\begin{equation}
\begin{array}{cc}
\thickmuskip=0mu
\medmuskip=0mu
\thinmuskip=0mu -\log \det\Delta=\int_{0}^{\infty}\frac{dt}{t}\int d^{3}x\sqrt{g}\sum_{n}\left\{g^{\mu\rho}\frac{\partial(\gamma^{n}x)^{\mu^{\prime}}}{\partial x^{\rho}}g^{\nu\sigma}\frac{\partial(\gamma^{n}x)^{\nu^{\prime}}}{\partial x^{\sigma}}\right.\nonumber\\
\left.\left(K_{\mu\mu^{\prime},\nu\nu^{\prime}}^{\mathbb{\mathbb{H}}}(t,r(x,\gamma^{n}x))+K_{\mu\mu^{\prime},\nu\nu^{\prime}}^{\mathbb{\mathbb{H}}}(t,r(x_{mirror},\gamma^{n}x))\right)\right\}.\end{array}
\end{equation}
Let us consider the calculation of the first term. The integral over
$r$ gives
\begin{equation}
\begin{array}{cc}\thickmuskip=0mu
\medmuskip=0mu
\thinmuskip=0mu 
\!\int_{0}^{\infty}\!\frac{dt}{t}\!\int d^{3}x\!\sqrt{g}\!\underset{n}{\sum}\!g^{\mu\rho}\!\frac{\partial(\gamma^{n}x)^{\mu^{\prime}}}{\partial x^{\rho}}\!g^{\nu\sigma}\!\frac{\partial(\gamma^{n}x)^{\nu^{\prime}}}{\partial x^{\sigma}}\!K_{\!\mu\!\mu\!^{\prime}\!,\!\nu\!\nu\!^{\prime}}^{\mathbb{\mathbb{H}}}\!(t,r(x,\gamma^{n}x))\\
=\sum\!\int_{0}^{\infty}\!\frac{dt}{t}\!\frac{2\pi^{2}\tau_{2}}{2(\cosh\beta-\cos\alpha)}\!\frac{e^{-\frac{\beta^{2}}{4t}}}{2\pi^{\frac{3}{2}}\sqrt{t}}\!(e^{-t}\!\cos2\alpha\!+\!e^{-4t}\!\cos\alpha\!+\!\frac{e^{-5t}}{2})\!\\
=\sum_{n=1}^{\infty}\frac{1}{n(\cosh\beta-\cos\alpha)}(e^{-\beta}\cos2\alpha+e^{-2\beta}\cos\alpha+\frac{e^{-\sqrt{5}\beta}}{2}),
\end{array}
\end{equation}
where we omit an $n=0$ term because it is proportional to the volume.
Next, we consider the second term. We can perform the integration similarly:

\begin{eqnarray}
\thickmuskip=0mu
\medmuskip=0mu
\thinmuskip=0mu 
&&\int_{0}^{\infty}\frac{dt}{t}\int d^{3}x\sqrt{g}\underset{n}{\sum}\left\{g^{\mu\rho}\frac{\partial(\gamma^{n}x)^{\mu^{\prime}}}{\partial x^{\rho}}g^{\nu\sigma}\frac{\partial(\gamma^{n}x)^{\nu^{\prime}}}{\partial x^{\sigma}}\right.\nonumber\\
&&\left.K_{\mu\mu^{\prime},\nu\nu^{\prime}}^{\mathbb{\mathbb{H}}}(t,r(x_{mirror},\gamma^{n}x))\right\}\nonumber\\
&&=\sum_{n}\int_{0}^{\infty}\frac{dt}{t}2\pi\tau_{2}\int_{-\frac{\pi}{2}}^{\frac{\pi}{2}}d\phi\left\{\frac{1}{2(\cosh\beta+\cos(2\phi-\alpha))}\right.\nonumber\\
&&\left.\frac{e^{-\frac{\beta^{2}}{4t}}}{2\pi^{\frac{3}{2}}\sqrt{t}}(e^{-t}\cos2\alpha+e^{-4t}\cos\alpha+\frac{e^{-5t}}{2})\right\}\nonumber\\
&=&\!\int_{0}^{\infty}\!\frac{dt}{t}\!2\pi\tau_{2}\!\int_{-\frac{\pi}{2}}^{\frac{\pi}{2}}\!d\phi\!\frac{1}{2(1+\cos(2\phi))}\!\frac{1}{2\pi^{\frac{3}{2}}\sqrt{t}}\!(e^{-t}\!+\!e^{-4t}\!+\!\frac{e^{-5t}}{2}\!)\nonumber\\
&+&\sum_{n}\!\int_{0}^{\infty}\!\frac{dt}{t}\!
2\!\pi\!\tau_{2}\!\frac{\pi}{\sinh\beta}\!\frac{\!e^{-\frac{\beta^{2}}{4t}}}{2\pi^{\frac{3}{2}}\sqrt{t}}\!(e^{-t}\!\cos2\alpha\!+\!e^{-4t}\cos\alpha\!+\!\frac{e^{-5t}}{2}\!).\nonumber\\
\end{eqnarray}

The first term contains a divergent integral and here we will ignore
it. The second term can be simplified as follows:
\begin{equation}
\begin{array}{cc}
\sum_{n}\int_{0}^{\infty}\frac{dt}{t}2\pi\tau_{2}\frac{\pi}{\sinh\beta}\frac{e^{-\frac{\beta^{2}}{4t}}}{2\pi^{\frac{3}{2}}\sqrt{t}}(e^{-t}\cos2\alpha+e^{-4t}\cos\alpha+\frac{e^{-5t}}{2})\\
=\sum_{n=1}^{\infty}\frac{1}{n\sinh\beta}(e^{-\beta}\cos2\alpha+e^{-2\beta}\cos\alpha+\frac{e^{-\sqrt{5}\beta}}{2}).
\end{array}
\end{equation}

\subsection{A one-loop partition function for gravity }

We will consider a linearized graviton perturbation $h_{\mu\nu}$
around the AdS background $g_{\mu\nu}$. The Einstein-Hilbert action
in the three dimensions with a negative cosmological constant is given
by
\begin{equation}
S_{GR}=-\frac{1}{16\pi G}\int d^{3}x(R+2)\sqrt{g}.\label{eq:}
\end{equation}
We will use the gauge of \citep{tHooft:1974toh}, where we add the
gauge-fixing term to (\ref{eq:}):
\begin{equation}
S_{GF}=\frac{1}{32\pi G}\int d^{3}x\sqrt{g}\nabla^{\mu}(h_{\mu\sigma}-\frac{1}{2}g_{\mu\sigma}h)\nabla^{\nu}(h_{\nu}^{\sigma}-\frac{1}{2}\delta_{\nu}^{\sigma}h).
\end{equation}
It is convenient to define the traceless part and the trace part:
\begin{equation}
\begin{array}{cc}
\phi_{\mu\nu}=h_{\mu\nu}-\frac{1}{3}g_{\mu\nu}h_{\rho}^{\rho}\\
\phi=h_{\rho}^{\rho}.
\end{array}
\end{equation}
The gauge-fixed action is given by \citep{Giombi:2008vd}
\begin{eqnarray}
S&=&-\frac{1}{32\pi G}\int d^{3}x\sqrt{g}\left\{ \frac{1}{2}\phi_{\mu\nu}(g^{\mu\rho}g^{\nu\sigma}\nabla^{2}+2R^{\mu\rho\nu\sigma})\phi_{\rho\sigma}\right.\nonumber\\
&-&\left.\frac{1}{12}\phi(\nabla^{2}-4)\phi\right\} .
\end{eqnarray}
We will wick-rotate $\phi\rightarrow i\phi$ in order to make the
kinetic term positive definite. The gauge-fixing term introduces a
Fadeev-Popov field, which is a Grassmann-odd vector field:

\begin{equation}
S_{ghost}=\frac{1}{32\pi G}\int d^{3}x\sqrt{g}\overline{\eta_{\mu}}(-g^{\mu\nu}\nabla^{2}-R^{\mu\nu})\eta^{\nu}.
\end{equation}
Therefore, the gravity partition function can be obtained by subtracting
the contribution of the vector ghost field with $m^{2}=4$ and the
scalar field with $m^{2}=4$, which corresponds to the trace part
of the fluctuation:
\begin{eqnarray}
&&\log Z_{gravity}^{1-loop}\nonumber\\
&=&-\frac{1}{2}\log\det\Delta^{graviton}+\log\det\Delta^{vector}-\frac{1}{2}\log\det\Delta^{scalar}\nonumber\\
&=&\sum_{n=2}^{\infty}-\frac{\log\left|1-q^{n}\right|^{2}}{2}\nonumber\\
&+&\sum_{n=1}^{\infty}\frac{1}{2n}\frac{1}{1-\left|q\right|^{2n}}(q^{2n}(1-\overline{q}^{n})+\overline{q}^{2n}(1-q^{n})).
\end{eqnarray}

The first term is just a half of the contribution in the original
theory and the second term comes from the mirror contribution.

\section{A physical Interpretation}

\subsection{A BCFT interpretation of the partition function}

In this section we summarize the results of the partition functions
in the previous section and give a physical interpretation from
the BCFT viewpoint. Before we move on, let us review our strategy of
the calculation. We calculated the partition function using the method
of images. In our calculation we approximate the ETW brane as the
hard wall in the bulk geometry, which means that the location
of the ETW brane is determined by solving the Neumann boundary condition for the tensionless ETW brane $K_{ab}=0$. Then, we apply
the method of images to the background AdS metric and evaluate the
fluctuation of the metric around the solution. 
Because the brane is tensionless and the action is proportional
to the tension, here we obtain the contributions from the ETW brane only
through the method of images.

Let us summarize our result. We firstly consider the scalar part.
The partition function of the free scalar field with Neumann boundary
condition on the ETW brane is given by
\begin{eqnarray}
Z_{Scalar}=\left(\prod_{l=0}^{\infty}\prod_{l^{\prime}=0}^{\infty}\frac{1}{\sqrt{1-q^{l+h}\overline{q}^{l^{\prime}+h}}}\right)\nonumber\\
\cdot\left(\prod_{m=0}^{\infty}\frac{1}{\sqrt{1-q^{m+h}\overline{q}^{m+h}}}\right).
\end{eqnarray}

The first term describes the ordinary contribution 
and the second one does the mirror
image one. The former has a clear interpretation: it comes from a primary
field with conformal dimension $h$ and a summation over its descendants.
The square root is present because we take the volume of the space to be a
half of the original AdS space. The second term is coming from the
mirror image effect of the original field. However, because in the AdS/BCFT
case, the rotational symmetry of the torus is broken due to the ETW brane
, we have only one real parameter $\beta$. Therefore we expect that
we should set $q=\overline{q}$ physically. In this case the expression gets simplified:
\begin{eqnarray}
Z_{Scalar}=\left(\prod_{l=0}^{\infty}\prod_{l^{\prime}=0}^{\infty}\frac{1}{\sqrt{1-q^{l+l^{\prime}+2h}}}\right)\nonumber\\
\cdot\left(\prod_{m=0}^{\infty}\frac{1}{\sqrt{1-q^{2(m+h)}}}\right).
\end{eqnarray}

Next, the vector field partition function is expressed as follows:
\begin{eqnarray}
\thickmuskip=0mu
\medmuskip=0mu
\thinmuskip=0mu 
Z_{\!vector\!}\!=\!\left(\!\prod_{l,l^{\prime}=0}^{\infty}\!\frac{1}{\sqrt{\!1-q^{l+1}\overline{q}^{l^{\prime}}\!}\!\sqrt{\!1\!-\!q^{l}\overline{q}^{l^{\prime}+1}\!}}\!\frac{1}{\sqrt{\!1\!-\!q^{l+h}\overline{q}^{l^{\prime}+h}}\!}\!\right)\nonumber\\
\cdot\!\left(\!\prod_{[l=0}^{\infty}\!\frac{1}{\sqrt{1-q^{l+1}\overline{q}^{l}}\sqrt{1-q^{l}\overline{q}^{l+1}}}\!\frac{1}{\sqrt{1-q^{l+h}\overline{q}^{l+h}}}\!\right).\nonumber\\
\end{eqnarray}
As is the case before the first term describes the ordinary contribution 
and the second one does the mirror effect. 
In both lines, the second factor is due to the
contribution from the longitudinal scalar mode. 
The rest is coming from the transverse vector mode. The summation
is over $L_{-1}$ and $\overline{L}_{-1}$ plus descendants contribution,
which represents a massless spin-1 particle. For a massive vector field
we can replace the powers of $q$, i.e. $1+l$ with $1+l+h$. By setting
$q=\overline{q}$ as before for the consistent profile of ETW brane, 
the partition function takes the simplified form:
\begin{eqnarray}
Z_{vector}=\left(\!\prod_{l,l^{\prime}=0}^{\infty}\frac{1}{1-q^{l+l^{\prime}+1}}\frac{1}{\sqrt{1-q^{l+l^{\prime}+2h}}}\right)\nonumber\\
\cdot\left(\!\prod_{l=0}^{\infty}\frac{1}{1-q^{2l+1}}\frac{1}{\sqrt{1-q^{2(l+h)}}}\right).
\end{eqnarray}

Finally, let us consider the gravity partition function. After some algebras, we reach the expression
\begin{eqnarray}
Z_{gravity}&=&\left(\!\prod_{m=2}^{\infty}\frac{1}{\left|1-q^{m}\right|}\right)\nonumber\\
&&\cdot\left(\!\prod_{l=0}^{\infty}\frac{\sqrt{1-q^{l+2}\overline{q}^{l+1}}\sqrt{1-q^{l+1}\overline{q}^{l+2}}}{\sqrt{1-q^{l+2}\overline{q}^{l}}\sqrt{1-q^{l}\overline{q}^{l+2}}}\right).\nonumber\\\label{4.5}
\end{eqnarray}
The first part gives the summation over vacuum and its chiral Virasoro
descendants. The second term looks complicated, but it has important
physical meaning: the numerator represents the massive vector field
with $h=1$ and the denominator represents a massless spin-2 field.
In $AdS_{3}/CFT_{2}$, the bulk field with a spin $l$ and a mass
$M$ is dual to an operator with a conformal dimension $2h$:
\begin{equation}
\begin{array}{cc}
M^{2}=\Delta(\Delta-2)+l^{2}\\
\Delta=2h+l.
\end{array}
\end{equation}
Therefore, the $h=1$ massive vector field has a mass $M^{2}=4$ in
the bulk, which exactly matches what appears as a ghost vector field
when we fix the gauge redundancy of gravity. We will revisit this point
in sub-section 4-C. 

Finally, by setting $q=\bar{q}$ again for a consistent profile of the ETW brane, we obtain
\begin{equation}
Z_{gravity}=\!\prod_{l=0}^{\infty}\frac{1}{\left(1-q^{2l+2}\right)^{2}}.\label{4.7}
\end{equation}
This has a peculiar exponent $2l+2$ when compared with the original result (\ref{eq:-2}) without ETW branes \citep{Giombi:2008vd}. 
We note that even though taking
into the ghost contribution this partition function seems to be physical
in a sense that the coefficients of the expansion in powers of $q$, i,e, 
 the numbers of states at that conformal dimension, are non-negative.
Interestingly, the ghost contribution goes away.

\subsection{A one-loop partition function with the Dirichlet boundary condition}

So far we impose the Neumann boundary condition on the ETW brane, following the standard prescription of AdS/BCFT. 
However, it is also interesting to consider more general boundary conditions. Therefore, in this section we repeat the similar calculation now for the Dirichlet boundary condition on the ETW brane. Refer to 
\citep{Miao:2018qkc} which suggests a possibility of using the Dirichlet boundary condition in related holographic computations.
In this section we only present the results briefly, leaving the details
in the appendix B.

The procedure to find results in the Dirichlet case is very simple: just flip the sign of the mirror image contribution. This leads to
\begin{equation}
Z_{Scalar}=\frac{\left(\!\prod_{m=0}^{\infty}\frac{1}{\sqrt{1-q^{m+h}\overline{q}^{m+h}}}\right)}{\left(\!\prod_{l,l^{\prime}=0}^{\infty}\frac{1}{\sqrt{1-q^{l+h}\overline{q}^{l^{\prime}+h}}}\right)},
\end{equation}
\begin{equation}
Z_{vector}=\frac{\prod_{l=0}^{\infty}\frac{1}{\sqrt{1-q^{l+1}\overline{q}^{l}}\sqrt{1-q^{l}\overline{q}^{l+1}}}\frac{1}{\sqrt{1-q^{l+h}\overline{q}^{l+h}}}}{\prod_{l,l^{\prime}=0}^{\infty}\frac{1}{\sqrt{1-q^{l+1}\overline{q}^{l^{\prime}}}\sqrt{1-q^{l}\overline{q}^{l^{\prime}+1}}}\frac{1}{\sqrt{1-q^{l+h}\overline{q}^{l^{\prime}+h}}}},
\end{equation}
\begin{eqnarray}
Z_{gravity}&=&\left(\!\prod_{m=2}^{\infty}\frac{1}{\left|1-q^{m}\right|}\right)\nonumber\\
&&\cdot\left(\!\prod_{l=0}^{\infty}\frac{\sqrt{1-q^{l+2}\overline{q}^{l}}\sqrt{1-q^{l}\overline{q}^{l+2}}}{\sqrt{1-q^{l+2}\overline{q}^{l+1}}\sqrt{1-q^{l+1}\overline{q}^{l+2}}}\right).\nonumber\\
\end{eqnarray}

When we set $q=\overline{q}$ to have a consistent profile of ETW branes, we obtain
\begin{equation}
Z_{Scalar}=\!\prod_{l,l^{\prime}=0}^{\infty}\frac{1}{\sqrt{1-q^{l+l^{\prime}+2h}}},
\end{equation}

\begin{equation}
Z_{vector}^{transverse}=\!\prod_{l,l^{\prime}=0}^{\infty}\frac{1}{1-q^{l+l^{\prime}+1}},
\end{equation}

\begin{equation}
Z_{gravity}=\!\prod_{l=0}^{\infty}\frac{1}{(1-q^{2l+3})^{2}},\label{4.13}
\end{equation}
where we omit the longitudinal modes in the vector field partition
function. The multiples of non-diagonal contributions lead to the gravity partition function which includes only the odd integer modes unlike the Neumann case.
\subsection{A consistency of the boundary condition}

In this section we will be more careful about the boundary conditions.
Refer to \citep{vanNieuwenhuizen:2005kg,Vassilevich:2003xt,Witten:2018lgb,Moss:1996ip} for a list of early works.
Usually, in gravity we impose the Neumann or the Dirichlet boundary condition (B.C.) on a boundary. At the tree-level, these conditions harm nothing in
calculating physical quantities such as the partition function. 
However, the situation changes when we consider quantities 
at the one-loop level. At the linearized level of the metric, we need
a gauge-fixing term and a ghost field. The boundary conditions for
these fields affect the one-loop partition function and what is more,
the gauge transformation for the metric may not be compatible with
a given boundary condition. Thus, the boundary operator $B$ defines
a consistent gauge-invariant boundary condition
\begin{equation}
B\phi\mid_{\partial M}=0,
\end{equation}
if and only if there exist boundary conditions for the corresponding a 
gauge parameter $\xi$
\begin{equation}
B_{\xi}\xi=0,
\end{equation}
such that
\begin{equation}
B\delta_{\xi}\phi=0.
\end{equation}
This condition ensures the validity of the one-loop calculation of
the gauge invariance with the Faddeev-Popov trick. Now we revisit
our problem. In our calculation we impose the Neumann or the Dirichlet conditions
for all fields including ghost fields. Let us check if this is reasonable
or not. Firstly, we consider the Neumann case. Note that we now split
the metric into the background AdS Poincare metric $g_{\mu\nu}$ and
the metric fluctuation $h_{\mu\nu}$. We impose the Neumann B.C. on $h_{ij}$:
\begin{equation}
\partial_{x}h_{ij}\mid_{x=0}=0,
\end{equation}
where $i,j$ represents the tangential direction along the brane.
This comes from the original condition $K_{ab}=0.$ The symmetric
traceless tensor $\phi_{\mu\nu}$ and trace part $\phi$ are given
by
\begin{equation}
\begin{array}{cc}
\phi_{\mu\nu}=h_{\mu\nu}-\frac{1}{3}g_{\mu\nu}\phi,\\
\phi=h_{\rho}^{\rho}.
\end{array}
\end{equation}
Therefore, we should also impose the Neumann B.C. for these fields:
\begin{equation}
\begin{array}{cc}
\partial_{x}\phi_{ij}\mid_{x=0}=0,\\
\partial_{x}\phi\mid_{x=0}=0.
\end{array}
\end{equation}
To see the condition for the ghost field we remember that it generates a 
gauge transformation for the metric:
\begin{equation}
h_{\mu\nu}\rightarrow h_{\mu\nu}+\nabla_{\mu}\eta_{\nu}+\nabla_{\nu}\eta_{\mu}.
\end{equation}
Let us check the gauge invariance of the Neumann B.C. The condition for
the ghost field is

\begin{equation}
\partial_{x}(\nabla_{i}\eta_{j}+\nabla_{j}\eta_{i})=0.
\end{equation}
Since the Christoffel symbols are functions of $y$, we can satisfy this
equation if we impose the Neumann B.C. for the ghost vector field $\eta_{i}$
and the anti-ghost vector field $\overline{\eta_{i}}$. 

Next, we move
to the components in the normal direction. If we allow the boundary
to fluctuate infinitesimally along the $x$-direction, then we should
impose the Neumann boundary condition for $\eta_{x}$ and $\overline{\eta_{x}}$:
\begin{equation}
\begin{array}{cc}
\partial_{x}\eta_{x}\mid_{x=0}=0,\\
\partial_{x}\overline{\eta_{x}}\mid_{x=0}=0.
\end{array}\label{4.22}
\end{equation}
Additionally, the BRST variation for $\overline{\eta_{\mu}}$ is given
by

\begin{equation}
\delta\overline{\eta_{\mu}}=\nabla^{\nu}h_{\mu\nu}-\frac{1}{2}\partial_{\mu}\phi.
\end{equation}
This variation should also satisfy (\ref{4.22}). Thus we can get
the boundary condition of the metric along the normal direction:

\begin{equation}
\partial_{x}\nabla_{\nu}h^{\mu\nu}\mid_{x=0}=0.
\end{equation}
Since the Christoffel symbols are only functions of $y$, we can satisfy
the above equation if we impose
\begin{equation}
\partial_{x}h_{\mu\nu}\mid_{x=0}=0.
\end{equation}
This is what we explicitly assumed in section 3. However, we can
consider one more possible boundary condition. If we strictly fix
the boundary to be on $x=0$, then we should impose the Dirichlet
boundary condition for $\eta_{x}$ and $\overline{\eta_{x}}$:

\begin{equation}
\begin{array}{cc}
\eta_{x}\mid_{x=0}=0,\\
\overline{\eta_{x}}\mid_{x=0}=0.
\end{array}\label{4.26}
\end{equation}
Correspondingly the BRST variation of the ghost field changes as
\begin{equation}
\delta\overline{\eta_{\mu}}=\nabla^{\nu}h_{\mu\nu}-\frac{1}{2}\partial_{\mu}\phi=0,
\end{equation}
at the boundary $x=0$. This equation determines the boundary condition
for the metric along $x$ direction.

Next, we consider the Dirichlet case. For the tangential direction
we should impose the Dirichlet B.C.
\begin{equation}
\begin{array}{cc}
h_{ij}=0,\\
\phi_{ij}=\phi=0,\\
\eta_{i}=\overline{\eta_{i}}=0,
\end{array}
\end{equation}
at the boundary $x=0$. If we require that the gauge transformation
should vanish at $x=0$, then we get

\begin{equation}
\begin{array}{cc}
\eta_{x}\mid_{x=0}=0,\\
\overline{\eta_{x}}\mid_{x=0}=0.
\end{array}
\end{equation}
The BRST variation gives additional constraints on the metric
\begin{equation}
\delta\overline{\eta_{\mu}}=\nabla^{\nu}h_{\mu\nu}-\frac{1}{2}\partial_{\mu}\phi=0.\label{4.30}
\end{equation}
This result is previously discussed in \citep{Witten:2018lgb} and
is somewhat remarkable; in the Neumann case we show that we can impose
the Neumann condition for all fields along the $x$-direction at the
boundary. However, in the Dirichlet case we must impose the Neumann like
boundary condition for the metric as in (\ref{4.30}). Therefore, our
calculation in section 3 in the Dirichlet case is not BRST invariant
because we previously imposed the Dirichlet boundary condition for 
all the fields and all the components at the boundary. For the
completion of classifying the boundary condition we consider one more
case. If we allow the boundary to fluctuate along the $x$ direction,
we get
\begin{equation}
\partial_{x}\eta_{x}=\partial_{x}\overline{\eta_{x}}=0,
\end{equation}
at the boundary. Then, BRST variation gives
\begin{equation}
\partial_{x}\left(\nabla_{\nu}h^{\mu\nu}-\frac{1}{2}\partial^{\nu}\phi\right)=0,
\end{equation}
at the boundary. However, in the Dirichlet case we should consider
one more problem: the ellipticity of the differential operators. In \citep{Witten:2018lgb}, the author discussed that Euclidean linearized gravity with purely the Dirichlet B.C. is not elliptic and hence perturbatively ill-defined.
We note that in the AdS/CFT case we can allow the Dirichlet B.C. up
to the Weyl transformation of the metric and this B.C. is elliptic
\citep{Witten:2018lgb}. As a consistency check let us consider
the ellipticity of the Neumann B.C. at the level of \citep{Witten:2018lgb}.
Now we impose the Neumann B.C. on all the components of the metric.
Then the metric can be written as
\begin{equation}
h_{\mu\nu}=\zeta_{\mu\nu}\cos\left(k_{x}x\right)\exp\left(ik_{i}x^{i}\right).
\end{equation}
If we differentiate with respect to the $x$, then at the boundary
it vanishes automatically. 

In the Dirichlet case, we did not notice any pathology in our
calculation of the partition function. However, we already know that
this boundary condition is not elliptic.  Indeed, we can see the breakdown of the ellipticity in a similar way to the above case. We can take the Fourier transformation of the kernel and
$x$ dependence comes from $\exp\left(ikx\right)$. Then, the above
argument shows that in the Dirichlet case the B.C. is not elliptic.
We expect that in our calculation the infinitely many zero-modes are
hidden in our regularization by the zeta function.



We summarize the above discussion. Firstly in both the Neumann and
the Dirichlet case we have two sets of boundary conditions for the
normal component of the vector fields like in (\ref{4.22}) and (\ref{4.26}).
We found that our calculation in section 3 will be consistent with BRST
invariance for the Neumann boundary condition on all the fields and
all the components. On the other hand, if we impose the Dirichlet B.C. 
on all the components of the metric, then the differential
operator is not elliptic, and thus it is perturbatively ill-defined. 
However, in the Neumann case we can check that the differential operator is
elliptic at the level of \citep{Witten:2018lgb}.

\subsection{$SL(2,\mathbb{Z})$ summation of the partition function}

Since we derived the one-loop partition function in section 3, here we would like to consider taking $SL(2,\mathbb{Z})$ summation as in \citep{Dijkgraaf:2000fq,Maloney:2007ud}.
At first sight this seems to be hard because the first term in (\ref{4.5})
cannot be expanded as a polynomial of $q$ and $\overline{q}$. However,
physically we remember that we can set $q=\overline{q}$ , so we will
use (\ref{4.7}) and (\ref{4.13}).

Firstly, let us consider the Neumann case. In section 1 we calculated
the tree-level partition function of the BTZ black hole. To derive
the tree-level partition function of the thermal AdS, we can change
the modular parameter as $\tau\rightarrow-\frac{1}{\tau}$. Therefore,
partition function of the thermal AdS is given by
\begin{eqnarray}
Z_{0,1}(\tau)&=&\left|q\overline{q}\right|^{-\frac{k}{2}}\left(\!\prod_{m=2}^{\infty}\frac{1}{\left|1-q^{m}\right|}\right)\nonumber\\
&&\cdot\left(\!\prod_{l=0}^{\infty}\frac{\sqrt{1-q^{l+2}\overline{q}^{l+1}}\sqrt{1-q^{l+1}\overline{q}^{l+2}}}{\sqrt{1-q^{l+2}\overline{q}^{l}}\sqrt{1-q^{l}\overline{q}^{l+2}}}\right)\nonumber\\
&=&q^{-k}\prod_{l=0}^{\infty}\frac{1}{\left(1-q^{2l+2}\right)^{2}},
\end{eqnarray}
where $k=\frac{1}{16G}$. The whole classical Einstein solution with
the boundary torus is obtained by implementing the modular transformation:
\begin{equation}
\tau\rightarrow\gamma\tau=\frac{a\tau+b}{c\tau+d},
\end{equation}
where $\gamma\in SL(2,\mathbb{Z})/\{\pm1\}$. Hence, here we can take
$c>0$ and sum over $\left(c,d\right)$, which are relatively prime
integers. The full partition function can be written as
\begin{eqnarray}
Z(\tau)&=&\underset{(c,d)}{\sum}q^{-k}\prod_{l=0}^{\infty}\frac{1}{\left(1-q^{2l+2}\right)^{2}}\nonumber\\
&=&\underset{(c,d)}{\sum}\frac{q^{-k+\frac{1}{6}}}{\eta(2\tau)^{2}}\nonumber\\
&=&\frac{1}{\sqrt{\Im(2\tau)}\eta(2\tau)^{2}}\underset{(c,d)}{\sum}\left(\sqrt{\Im(2\tau)}q^{-k+\frac{1}{6}}\right)\mid_{\gamma}\nonumber\\
&=&\frac{E(\tau;\frac{k}{2}-\frac{1}{12},0)}{\sqrt{\Im(2\tau)}\eta(2\tau)^{2}},
\end{eqnarray}
where we define $E(\tau;n,m)=\underset{(c,d)}{\sum}(\sqrt{\Im(2\tau)}q^{n}\overline{q}^{m})$.
Here we use the modular invariance of $\sqrt{\Im(2\tau)}\eta(2\tau)^{2}$.

Next, we consider the Dirichlet case. As in the Neumann case, the
partition function becomes
\begin{eqnarray}
Z(\tau)&=&\underset{(c,d)}{\sum}q^{-k-\frac{1}{12}}\frac{\eta(2\tau)^{2}}{\eta(\tau)^{2}}(1-q)^{2}\nonumber\\
&=&\frac{\eta(2\tau)^{2}\sqrt{\Im(2\tau)}}{\eta(\tau)^{2}\sqrt{\Im(\tau)}}\underset{(c,d)}{\sum}(q^{-k-\frac{1}{12}}(1-q)^{2}\frac{\sqrt{\Im(\tau)}}{\sqrt{\Im(2\tau)}})\mid_{\gamma}.\nonumber\\
\end{eqnarray}
Though we can continue the calculation as in \citep{Maloney:2007ud},
we stop here and briefly discuss how to treat this partition function.
Firstly, this Poincare series is divergent, so we need some regularization.
One possible way is that we consider the following convergent series
\begin{equation}
\underset{(c,d)}{\sum}\left(\Im(2\tau)\right)^{s}q^{-k+\frac{1}{6}}.
\end{equation}
This series is convergent for $\Re(s)>1$. However, as is presented
in \citep{Maloney:2007ud}, we can take an analytic continuation
to $\Re(s)\leq1$ and especially at $s=\frac{1}{2}$ this series gets
regular. We expect that the spectrum has a negative density of states
as is the case in pure gravity \citep{Benjamin:2019stq,Benjamin:2020mfz}.
It will be an interesting future direction to specify the black-hole microstates using the modularity of the theory. 


\subsection{One-loop exactness of the partition function}

One-loop exactness of the partition function is an important problem
as in any other theory. Pure gravity in three dimensions is known to be one-loop exact because the bulk diffeomorphism is governed by the Virasoro symmetry, hence the partition function does not suffer any quantum correction other than the Virasoro descendants. We expect the same property in this study, though we have not explicitly shown it.
In the two-dimensional BCFT cases, we can use the double trick as explained in \citep{Polchinski:1998rq}, which leads to the identification
\begin{equation}
L_{n}=\overline{L_{-n}}.
\end{equation}
Therefore, a half of the Virasoro symmetry or only the chiral mode
survives. This also guarantees that if we can properly calculate, respecting
the BRST invariance, then the partition function of our case will be one-loop exact.

\section{Conclusions and Discussions}

In this paper we studied the partition function in the three dimensional AdS/BCFT
model with one loop quantum corrections.
First we calculated the tree-level partition function of the BTZ black hole
in the presence of the end of the world brane with arbitrary tension.
The result matches with the previous results in the non-rotating case. 
At the one-loop level, we calculated the thermal AdS partition function with
a tensionless brane. We found that the result depends heavily on the choice of
sets of boundary conditions. We explicitly calculated the partition
function in the case where all the components of all the fields satisfy
the Neumann boundary condition and also in another case where all the components
of all the fields satisfy the Dirichlet boundary condition. In the
Neumann case, we expect that the system will be consistent with the BRST
quantization. We found that the partition function actually contains remnants
of ghost fields. Note that this ghost mode does not arise from a wrong
sign of kinetic terms but does from a fermionic spin one field of the BRST ghost. However if we consider the physically sensible profile of end of the world branes, we need to set $q=\overline{q}$. This leads to a partition function with a healthy spectrum such that the number of states at each energy level is non-negative.
This suggests that the AdS/BCFT formulation with the Neumann boundary condition is consistent at the one-loop level.

In the Dirichlet case, on the other hand, we also encounter unphysical
modes. There are other possible sets of boundary conditions, namely, mixed boundary conditions for each component. However, in this
paper we have not calculated them due to its computational difficulty.
This will be one of our future problems.  It is also possible that there might be sensible models of the gravity with boundaries by adding a
suitable set of matter fields. For example, refer to \citep{vanNieuwenhuizen:2005kg}
for an argument using supersymmetry transformations.
Therefore, it will be interesting to specify the minimal set of boundary
conditions and matter fields to make the theory well-defined. For another
future direction it will be important to specify what kind
of mode localizes on the brane or decouples from the bulk. As another route to gravity, it may be interesting to understand our
problem in the Chern-Simons formulation. We expect that in this setup
we can understand the edge modes more clearly because we can only consider
the boundary conditions for gauge fields.

In order to derive the full partition function, we should sum over all possible
geometries with a given boundary condition \citep{Maldacena:2004rf}.
In this work, the bulk geometry is a half of the solid torus, therefore
we can sum over the torus moduli $SL(2,\mathbb{Z})$. As we discussed
briefly in section 4, the computation is similar to the one in \citep{Maloney:2007ud}. It will be an interesting direction
to develop this calculation further and analyze the spectrum using modular bootstrap in BCFTs as in the pure gravity case \citep{Benjamin:2020mfz,Benjamin:2019stq}.

\begin{acknowledgements}
We thank T.Takayanagi for correcting some errors in manuscripts and
encouraging me. We are also grateful to T.Nishioka, T.Ugajin, and
D.Vassilevich for valuable comments and discussions. 
\end{acknowledgements}

\appendix

\section*{Appendix A: Derivatives of a geodesic distance}

Here we present the calculation of the derivatives of $u$. Ordinary contributions are given as follows:

\begin{equation}
\begin{array}{cc}
\partial_{y}\partial_{x^{\prime}}u=-\frac{\cos(\theta)(e^{\beta}\cos(\phi-\alpha)-\cos(\phi))}{\rho^{2}e^{\beta}\sin^{3}(\theta)},\\
\partial_{y}\partial_{\eta^{\prime}}u=-\frac{\cos(\theta)(e^{\beta}\sin(\phi-\alpha)-\sin(\phi))}{\rho^{2}e^{\beta}\sin^{3}(\theta)},\\
\partial_{x}\partial_{y^{\prime}}u=-\frac{\cos(\theta)(-e^{\beta}\cos(\phi-\alpha)+\cos(\phi))}{\rho^{2}e^{2\beta}\sin^{3}(\theta)},\\
\partial_{\eta}\partial_{y^{\prime}}u=-\frac{\cos(\theta)(-e^{\beta}\sin(\phi-\alpha)+\sin(\phi))}{\rho^{2}e^{2\beta}\sin^{3}(\theta)},\\
\partial_{y}\partial_{y^{\prime}}u=-\frac{(2\cosh\beta-\cosh r)}{\rho^{2}e^{\beta}\sin^{2}(\theta)},\\
\partial_{x}\partial_{x^{\prime}}u=-\frac{1}{\rho^{2}e^{\beta}\sin^{2}(\theta)},\\
\partial_{\eta}\partial_{\eta^{\prime}}u=-\frac{1}{\rho^{2}e^{\beta}\sin^{2}(\theta)},\\
\partial_{x}\partial_{\eta^{\prime}}u=\partial_{\eta}\partial_{x^{\prime}}u=0,\\
\partial_{y}u=\frac{1}{\rho\sin\theta}(e^{-\beta}-\cosh r),\\
\partial_{x}u=-\frac{\cos(\theta)(e^{\beta}\cos(\phi-\alpha)-\cos(\phi))}{\rho e^{\beta}\sin^{2}(\theta)},\\
\partial_{\eta}u=-\frac{\cos(\theta)(e^{\beta}\sin(\phi-\alpha)-\sin(\phi))}{\rho e^{\beta}\sin^{2}(\theta)},\\
\partial_{y^{\prime}}u=\frac{1}{\rho\sin\theta e^{\beta}}(e^{\beta}-\cosh r),\\
\partial_{x^{\prime}}u=\frac{\cos(\theta)(e^{\beta}\cos(\phi-\alpha)-\cos(\phi))}{\rho e^{\beta}\sin^{2}(\theta)},\\
\partial_{\eta^{\prime}}u=\frac{\cos(\theta)(e^{\beta}\sin(\phi-\alpha)-\sin(\phi))}{\rho e^{\beta}\sin^{2}(\theta)}.
\end{array}
\end{equation}

The mirror image contributions are given by
\begin{equation}
\begin{array}{cc}
\partial_{y}\partial_{x^{\prime}}u=-\frac{\cos(\theta)(e^{\beta}\cos(\phi-\alpha)+\cos(\phi))}{\rho^{2}e^{\beta}\sin^{3}(\theta)},\\
\partial_{y}\partial_{\eta^{\prime}}u=-\frac{\cos(\theta)(e^{\beta}\sin(\phi-\alpha)-\sin(\phi))}{\rho^{2}e^{\beta}\sin^{3}(\theta)},\\
\partial_{x}\partial_{y^{\prime}}u=\frac{\cos(\theta)(e^{\beta}\cos(\phi-\alpha)+\cos(\phi))}{\rho^{2}e^{2\beta}\sin^{3}(\theta)},\\
\partial_{\eta}\partial_{y^{\prime}}u=-\frac{\cos(\theta)(-e^{\beta}\sin(\phi-\alpha)+\sin(\phi))}{\rho^{2}e^{2\beta}\sin^{3}(\theta)},\\
\partial_{y}\partial_{y^{\prime}}u=-\frac{(2\cosh\beta-\cosh r)}{\rho^{2}e^{\beta}\sin^{2}(\theta)},\\
\partial_{x}\partial_{x^{\prime}}u=-\frac{1}{\rho^{2}e^{\beta}\sin^{2}(\theta)},\\
\partial_{\eta}\partial_{\eta^{\prime}}u=-\frac{1}{\rho^{2}e^{\beta}\sin^{2}(\theta)},\\
\partial_{x}\partial_{\eta^{\prime}}u=\partial_{\eta}\partial_{x^{\prime}}u=0,\\
\partial_{y}u=\frac{1}{\rho\sin\theta}(e^{-\beta}-\cosh r),\\
\partial_{x}u=-\frac{\cos(\theta)(e^{\beta}\cos(\phi-\alpha)+\cos(\phi))}{\rho e^{\beta}\sin^{2}(\theta)},\\
\partial_{\eta}u=-\frac{\cos(\theta)(e^{\beta}\sin(\phi-\alpha)-\sin(\phi))}{\rho e^{\beta}\sin^{2}(\theta)},\\
\partial_{y^{\prime}}u=\frac{1}{\rho\sin\theta e^{\beta}}(e^{\beta}-\cosh r),\\
\partial_{x^{\prime}}u=\frac{\cos(\theta)(e^{\beta}\cos(\phi-\alpha)+\cos(\phi))}{\rho e^{\beta}\sin^{2}(\theta)},\\
\partial_{\eta^{\prime}}u=\frac{\cos(\theta)(e^{\beta}\sin(\phi-\alpha)-\sin(\phi))}{\rho e^{\beta}\sin^{2}(\theta)}.
\end{array}
\end{equation}

\section*{Appendix B: Calculation in the Dirichlet case}

Let us consider the case where we impose the Dirichlet B.C. on all
the components of all the fields. In this case we can flip the sign
of the mirror image contributions in the calculation of the heat kernel. For the
scalar part, we get

\begin{eqnarray}
S_{scalar}^{(1)}&=&\frac{1}{2}\int_{+0}^{\infty}\frac{dt}{t}\int_{thermal\,AdS}d^{3}x\underset{n}{\sum}\sqrt{g}\nonumber\\
&&\cdot\left(K^{\mathbb{\mathbb{H}}}(t,x,\gamma^{n}x)-K^{\mathbb{\mathbb{H}}}(t,x^{mirror},\gamma^{n}x)\right)\nonumber\\
&=&\sum_{n=1}^{\infty}\frac{\left|q\right|^{n\left(1+\sqrt{m^{2}+1}\right)}}{n\left|1-q^{n}\right|^{2}}-\sum_{n=1}^{\infty}\frac{\left|q\right|^{n\left(1+\sqrt{m^{2}+1}\right)}}{n\left(1-\left|q\right|^{2n}\right)}.
\end{eqnarray}
The vector and symmetric traceless part can also be calculated
\begin{eqnarray}
S_{vector}^{(1)}&=&\int_{0}^{\infty}\frac{dt}{t}\int d^{3}x\sqrt{g}\underset{n}{\sum}g^{\mu\rho}\frac{\partial(\gamma^{n}x)^{\nu^{\prime}}}{\partial x^{\rho}}\nonumber\\
&&\cdot\left(K_{\mu\nu^{\prime}}^{\mathbb{\mathbb{H}}}(t,r(x,\gamma^{n}x))-K_{\mu\nu^{\prime}}^{\mathbb{\mathbb{H}}}(t,r(x_{mirror},\gamma^{n}x))\right)\nonumber\\
&=&\sum_{n=1}^{\infty}\frac{q^{n}+\overline{q}^{n}+\left|q\right|^{2n}}{n\left|1-q^{n}\right|^{2}}-\sum_{n=1}^{\infty}\frac{q^{n}+\overline{q}^{n}+\left|q\right|^{2n}}{n(1-\left|q\right|^{2n})},
\end{eqnarray}

and
\begin{eqnarray}
S_{spin-2}^{(1)}&=&\int_{0}^{\infty}\frac{dt}{t}\int d^{3}x\sqrt{g}\underset{n}{\sum}g^{\mu\rho}\frac{\partial(\gamma^{n}x)^{\mu^{\prime}}}{\partial x^{\rho}}g^{\nu\sigma}\frac{\partial(\gamma^{n}x)^{\nu^{\prime}}}{\partial x^{\sigma}}\nonumber\\
&&\cdot\left(\!K_{\mu\mu^{\prime},\nu\nu^{\prime}}^{\mathbb{\mathbb{H}}}(t,r(x,\gamma^{n}x))\!-\!K_{\mu\mu^{\prime},\nu\nu^{\prime}}^{\mathbb{\mathbb{H}}}\!(t,r(x_{mirror},\gamma^{n}x))\!\right)\nonumber
\\
&=&\sum_{n=1}^{\infty}\left(\frac{1}{n(\cosh\beta-\cos\alpha)}-\frac{1}{n\sinh\beta}\right)\nonumber\\
&&\cdot\left(e^{-\beta}\cos2\alpha+e^{-2\beta}\cos\alpha+\frac{e^{-\sqrt{5}\beta}}{2}\right).
\end{eqnarray}

Finally the gravity partition function becomes
\begin{eqnarray}
log Z_{gravity}^{1-loop}
&=&-\log\frac{\sqrt{\det\Delta^{graviton}\det\Delta^{scalar}}}{\det\Delta^{vector}}\nonumber\\
&=&-\sum_{n=2}^{\infty}\frac{\log\left|1-q^{n}\right|^{2}}{2}\nonumber\\
&-&\sum_{n=1}^{\infty}\frac{1}{2n}\frac{1}{1-\left|q\right|^{2n}}(q^{2n}(1-\overline{q}^{n})+\overline{q}^{2n}(1-q^{n})).\nonumber\\
\end{eqnarray}

\end{document}